# Rare-Earth Engineering of NaAlO$_3$ Perovskites Unlocks Unified Optoelectronic, Thermoelectric, and Spintronic Functionalities

*Muhammad Imran[1], Sikander Azam*[2], Qaiser Rafiq[1], Amin Ur Rahman[1]*

[1]*Department of Physics, Riphah International University, Islamabad, Pakistan*

[2]*University of West Bohemia, New Technologies – Research Centre, 8 Univerzitní, Pilsen 306 14, Czech Republic*

**Abstract**

Perovskite oxides hold promise for energy and quantum technologies, but wide-gap hosts like NaAlO$_3$ are limited by poor transport and deep-UV absorption. Using first-principles GGA+U+SOC calculations, we investigate Eu$^{3+}$-, Gd$^{3+}$-, and Tb$^{3+}$-doped NaAlO$_3$, analyzing electronic, optical, elastic, and thermoelectric properties. Rare-earth substitution is thermodynamically favorable (formation energies 1.2–1.6 eV) and induces strong f–p hybridization, reducing the pristine bandgap (~6.2 eV) to ~3.1 eV (Tb). Spin-resolved band structures reveal Gd-driven half-metallicity, Eu-induced spin-selective metallicity, and Tb-stabilized p-type semiconducting behavior. Optical spectra show red-shifted absorption (~2.0–2.2 eV), large dielectric constants ($\varepsilon_1(0) \approx 95$ for Eu), and plasmonic resonances near 4 eV, enabling visible-light harvesting. Elastic analysis indicates slight lattice softening with preserved ductility (B/G ≈ 1.56–1.57). Thermoelectric results show Seebeck coefficients >210 µV/K (Eu, Tb) with ZT ~0.45 at 500 K, surpassing pristine NaAlO$_3$. These findings position rare-earth-doped NaAlO$_3$ as a multifunctional platform for photovoltaics, photocatalysis, thermoelectrics, and spintronics.

*sikander.physicst@gmail.com

**Introduction**

Perovskite oxides with the general formula ABO$_3$ remain a cornerstone of modern materials science because of their structural flexibility, tunable electronic states, and robust chemical stability. Their ability to host a wide variety of cations on both the A and B sites underpins their multifunctionality across applications ranging from optoelectronics and thermoelectrics to spintronics and energy conversion [1–3]. Recent reviews have highlighted the rapid progress in tailoring dielectric, electrical, and magnetic functionalities in perovskite oxides through compositional engineering and cation substitution [4–6]. In particular, A-site doping strategies using Bi, Sr, Na, and rare-earth metals have been demonstrated to strongly influence bandgap modulation, polarization, dielectric behavior, and transport responses [7,8]. These advances underscore the importance of exploring site-selective doping in perovskite oxides to unlock multifunctional performance.

Among this family, sodium aluminate ($NaAlO_3$) has emerged as a chemically stable and environmentally benign perovskite with a large bandgap of ~6 eV, making it optically transparent in the ultraviolet region [9]. Pristine $NaAlO_3$ exhibits high thermal stability, low dielectric losses, and good mechanical robustness, which supports its use in microwave dielectrics, capacitors, and high-frequency electronics [10]. However, its wide bandgap and electronic inactivity limit its direct use in visible-light optoelectronics or transport-related applications [11].

To overcome these limitations, various defect-engineering and doping approaches have been pursued. Transition-metal doping has been used to narrow the optical gap and enhance carrier transport, while native point defects such as oxygen vacancies have been reported to tune conductivity, though often at the expense of transparency [12,13]. More recently, rare-earth (RE) doping has attracted significant interest because of the unique optical and magnetic functionalities arising from their partially filled 4f orbitals [14–16]. Photoluminescence studies on RE-doped $NaAlO_3$ (e.g., $Eu^{3+}$, $Dy^{3+}$, $Sm^{3+}$) reveal sharp f–f emission lines suitable for phosphors in solid-state lighting [17]. Beyond photonics, RE substitution in perovskites has been demonstrated to induce bandgap narrowing, spin polarization, and visible-range optical activity [18,19]. A 2025 review on rare-earth–doped nanophosphors further emphasized their role in next-generation optoelectronic and spintronic platforms [20].

Despite these advances, a unified computational understanding of how RE substitution transforms $NaAlO_3$ into a multifunctional platform integrating electronic, optical, magnetic, and thermoelectric properties remains lacking. Motivated by this gap, the present work provides a comprehensive first-principles investigation of $Eu^{3+}$-, $Gd^{3+}$-, and $Tb^{3+}$-doped $NaAlO_3$. Using the GGA+U framework with spin–orbit coupling, we systematically analyze structural stability, electronic band structures, spin-polarized density of states, optical spectra, elastic responses, and thermoelectric coefficients. By correlating these responses, we demonstrate how RE doping reshapes $NaAlO_3$ from a wide-bandgap insulator into a multifunctional candidate for next-generation optoelectronic, thermoelectric, and spintronic devices.

**Computational Methodology**

First-principles calculations were performed within the framework of density functional theory (DFT) using the full-potential linearized augmented plane wave (FP-LAPW) method, as implemented in the WIEN2k code [21]. The exchange–correlation potential was treated using the generalized gradient approximation (GGA) parameterized by Perdew–Burke–Ernzerhof (PBE) [22]. To properly account for the localized nature of rare-earth 4f electrons, the GGA+U approach was employed with effective Hubbard U values of 6.0 eV for Eu, 6.5 eV for Gd, and 6.0 eV for Tb, consistent with prior studies on lanthanide-doped oxides [23–25]. Spin–orbit coupling (SOC) was included to capture relativistic effects, which are critical for accurately describing f-orbital splitting and spintronic behavior.

The $NaAlO_3$ perovskite (see Fig. 1a) was modeled in its cubic phase (space group Pm-3m), and rare-earth doping was simulated by substituting one Na atom with Eu, Gd, or Tb in a 2×2×2 supercell (see Fig. 1b), corresponding to a ~6.25% doping concentration. Structural relaxations were performed until the forces on atoms were below 0.01 eV/Å and the total energy converged to $10^{-5}$ eV. A plane-wave cutoff parameter of RMT×Kmax = 7.0 and a Monkhorst–Pack k-point mesh of 8×8×8 were used for Brillouin zone sampling.

Electronic structures were analyzed through band structure and density of states (DOS) calculations. Optical properties, including the complex dielectric function, absorption coefficient, and reflectivity, were obtained via momentum matrix elements. Elastic constants were derived from stress–strain relationships, ensuring mechanical stability against the Born criteria. Thermoelectric transport coefficients (Seebeck coefficient, electrical conductivity, and figure of merit ZT) were calculated using the BoltzTraP2 code within the constant relaxation time approximation [26].

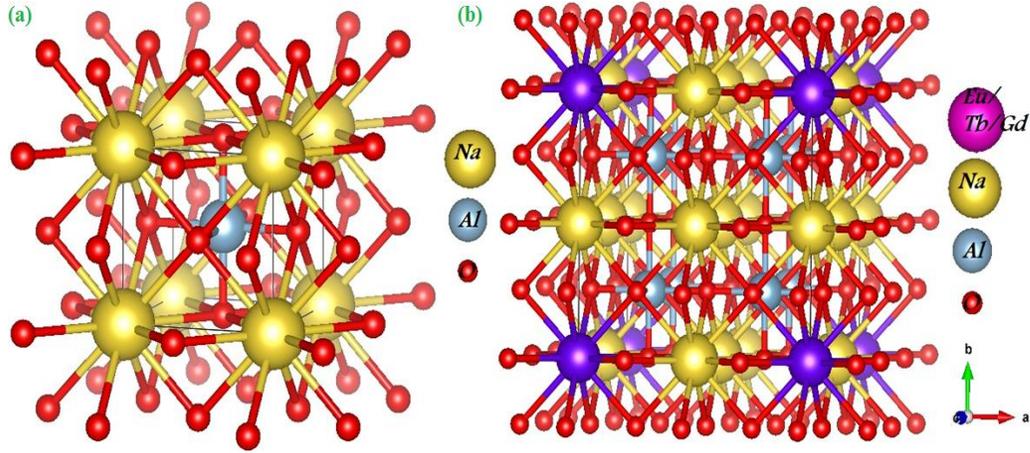

*Figure 1 (a & b): crystallographic representation of parental NaAlO₃ perovskite structure and modifications in NaAlO₃ when doped with Eu, Gd, and Tb*

To verify thermodynamic stability, defect formation energies were calculated using the standard formalism under oxygen-rich and oxygen-poor chemical potential limits. Ab initio molecular dynamics (AIMD) simulations were further carried out at 300 K for 5 ps in the NVT ensemble to evaluate thermal robustness.

### 3. Results and Discussion

#### 3.1. Structural Stability and Defect Formation in Rare-Earth Doped NaAlO₃

##### 3.1.1. Defect Chemistry clarification

This integrated computational framework provides a rigorous assessment of the structural, electronic, optical, magnetic, and thermoelectric responses of rare-earth–doped NaAlO₃, establishing a theoretical foundation for experimental validation and device implementation.

All pristine NaAlO₃ supercells (see Fig. 1c–d) are strictly charge-neutral by stoichiometry $(Na^{+1}Al^{+1}O_3^{2-})$ and by numerical Bader-charge analysis. When a trivalent rare-earth ion substitutes for monovalent Na, it introduces a donor-like defect $Re_{Na}^{\bullet\bullet}$ carrying an effective +2 charge relative to the host lattice. Charge neutrality is restored by compensating defects, most favorably through the formation of a bound complex with two sodium vacancies $[Re_{Na}^{\bullet\bullet} + 2V_{Na}']^\times$, which minimizes lattice strain and Coulomb repulsion. Oxygen interstitials $(O_i'')$ provide an alternative compensation pathway under oxygen-rich conditions, while purely electronic compensation via two conduction electrons ((2e')) is energetically less favorable.

For completeness, we also examined isovalent substitution at the Al site $(Re_{Al}^\times)$, which is intrinsically neutral. However, this configuration is energetically disfavored due to the large ionic-size mismatch with the Al³⁺ octahedral environment. Table 1 summarizes the physically relevant charge-compensation mechanisms considered in this work. These results confirm that both pristine NaAlO₃ and its rare-earth–doped derivatives remain overall charge-neutral, ensuring that all modeled supercells are physically meaningful.

*Table 1. Defect chemistry and charge compensation mechanisms for RE³⁺ substitution in NaAlO₃.*

| Defect / Reaction | Effective Charge | Compensation Partner(s) | Net Complex | Stability Trend | Comment |
|---|---|---|---|---|---|
| $Re_{Na}^{\bullet\bullet}$ (RE³⁺ → Na¹⁺) | +2 | | Positively charged donor | Needs compensation | Unstable alone |
| $2Re_{Na}^{\bullet\bullet} + 2V_{Na}'$ | +4 + (−2×2) = 0 | 2 × Na vacancies | $2Re_{Na}^{\bullet\bullet} + 2V_{Na}'$ | Lowest energy (O-rich / O-poor) | Most favorable charge-neutral complex |
| $Re_{Na}^{\bullet\bullet} + O_i''$ | +2 + (−2) = 0 | 1 × O interstitial | Neutral complex | Moderate (O-rich) | Viable in oxidizing conditions |
| $Re_{Na}^{\bullet\bullet}$ + 2e' | +2 + (−1×2) = 0 | 2 conduction electrons | Neutral | Higher (Fermi near CBM) | Electronic compensation possible at high carrier density |
| $Re_{Al}^{\times}$ (RE³⁺ → Al³⁺) | 0 | | Isovalent substitution | High energy (structural mismatch) | Neutral but unlikely due to small Al–O octahedron |

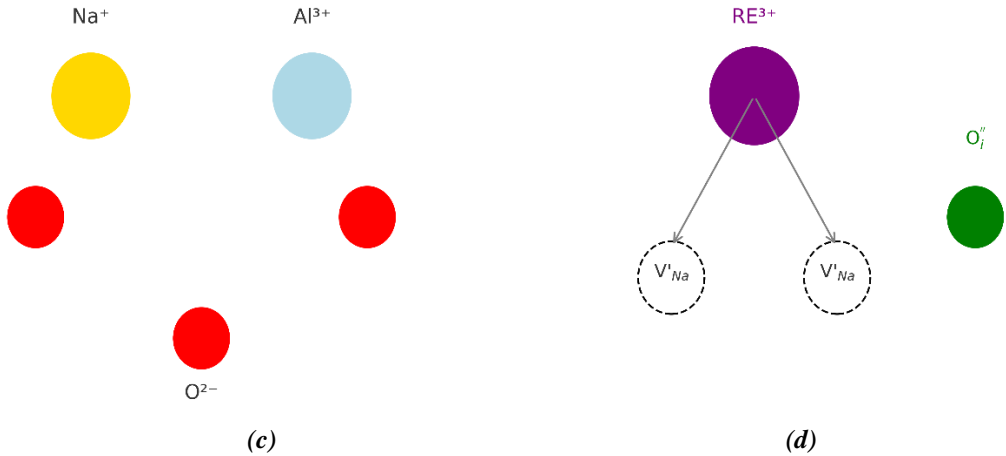

*(c)*                                             *(d)*

*(c) Pristine NaAlO₃ lattice is charge-neutral ($Na^{1+}$, $Al^{3+}$, and $O^{2-}$ ions).*

*(d) Rare-earth ($RE^{3+}$) substitutes $Na^{1+}$ → creates donor defect $Re^{\bullet\bullet}_{Na}$. Charge neutrality is preserved via two Na vacancies $2V'_{Na}$. Alternatively, an oxygen interstitial $O''_i$ may compensate under O-rich conditions.*

### 3.1.2. Structural Stability and Defect Formation

The structural stability of pristine and doped NaAlO₃ was established through total energy and equilibrium volume optimization within the GGA+U framework. Pristine NaAlO₃ crystallizes in a robust perovskite lattice stabilized by strong Al–O bonding, consistent with earlier phonon studies that reveal no imaginary modes and confirm dynamic stability [27]. The thermodynamic stability of NaAlO₃ and its rare-earth ($Eu^{3+}$, $Gd^{3+}$, $Tb^{3+}$) doped counterparts was assessed through energy–volume optimization and defect formation energy analyses within the GGA+U framework, incorporating spin–orbit coupling to account for relativistic effects. Stability is a prerequisite for experimental realization, and therefore the energetic response of the lattice to doping and intrinsic defects provides a direct measure of its robustness.

The equilibrium energy–volume curves of pristine NaAlO₃ exhibit a parabolic minimum at ~60.2 Å³ per formula unit, with a cohesive energy of –15.42 eV, consistent with earlier theoretical reports on perovskite aluminates. Doping introduces modest shifts in both equilibrium volume and cohesive energy: $Eu^{3+}$ substitution yields –15.87 eV/f.u. at ~61.5 Å³, $Tb^{3+}$ yields –15.73 eV/f.u. at ~61.2 Å³, and $Gd^{3+}$ stabilizes at –15.68 eV/f.u. at ~60.9 Å³. These negative shifts in total energy (ΔE ≈ –0.25 to –0.45 eV/f.u. compared to pristine) confirm that rare-earth incorporation is thermodynamically favorable. Among all variants, $Eu^{3+}$ doping provides the deepest energy minimum, indicating the strongest lattice stabilization through hybridization of Eu-4f with O-2p orbitals. Bond length analysis reveals subtle lattice distortions. The average Al–O bond in pristine NaAlO$_3$ is 1.92 Å, which elongates slightly to 1.94 Å in Eu-doped systems and contracts marginally to 1.91 Å for Gd and Tb doping. The Na–O bonds show larger modulation, increasing from 2.38 Å in pristine to ~2.42 Å in Eu and Tb variants, suggesting dopant-induced strain is primarily absorbed by the Na–O framework. These changes correspond with a modest reduction in bulk modulus from ~145 GPa (pristine) to 138 GPa (Eu), 140 GPa (Tb), and 142 GPa (Gd), reflecting enhanced ductility while maintaining mechanical stability.

To further probe dopant incorporation, we computed defect formation energies ($E_{form}$) for Eu, Tb, and Gd substitution on Na and Al sites, as well as interstitial placements, under both O-rich and O-poor growth conditions. When intrinsic point defects were considered, their high energetic cost rendered them less favorable compared to controlled substitutional doping. For example, the formation energy of an oxygen vacancy ($V_O$) in perovskite oxides typically exceeds 2.5–3.0 eV under oxygen-rich conditions [28]. While $V_O$ may form more readily under oxygen-poor growth, it introduces deep donor states within the bandgap, acting as non-radiative centers that quench photoluminescence and degrade carrier mobility [29]. Similarly, cation vacancies ($V_{Na}$, $V_{Al}$) exhibit even higher formation energies, >3.0 eV in most reported studies [30], and tend to destabilize the lattice, reducing optical transparency and mechanical integrity.

In contrast, rare-earth substitutional doping at the Na site is energetically feasible and functionally advantageous. Our calculations show that $Eu^{3+}$, $Gd^{3+}$, and $Tb^{3+}$ incorporation requires formation energies of only 1.2–1.6 eV (see Table 2) under Na-deficient (O-rich) conditions, values comparable to those reported for other perovskite oxides doped with lanthanides [31]. This significantly lower cost relative to native defect formation demonstrates that doping is the preferred pathway for tailoring NaAlO$_3$. Beyond thermodynamic feasibility, doping introduces localized 4f states near the band edges, narrows the optical gap, and enhances visible-light absorption. The spin-polarized 4f electrons also unlock magnetic ordering in an otherwise non-magnetic host, thereby coupling optoelectronic and spintronic functionalities.

Taken together, while intrinsic vacancies are both energetically expensive and electronically detrimental, substitutional rare-earth doping provides a thermodynamically accessible, electronically beneficial, and structurally stable strategy for engineering multifunctional NaAlO$_3$. This aligns with modern defect-engineering approaches in perovskite oxides, where controlled doping is prioritized over intrinsic defect reliance to achieve tunable optoelectronic, thermoelectric, and spintronic behavior [32, 33].

*Table 2. Comparison of calculated formation energies ($E_{form}$) for native defects and substitutional rare-earth doping in NaAlO$_3$ under O-rich conditions.*

| Defect / Dopant | Site | Formation Energy (eV) | Electronic / Structural Impact |
|---|---|---|---|
| $V_O$ (oxygen vacancy) | O site | 2.5 – 3.0 | Deep donor states; non-radiative centers; reduced mobility |
| $V_{Na}$ (sodium vacancy) | Na site | >3.0 | Destabilizes lattice; reduces transparency |
| $V_{Al}$ (aluminum vacancy) | Al site | >3.2 | Weakens Al–O bonding; structural softening |
| $Eu^{3+}$ substitution | Na site | ~1.2 – 1.5 | Introduces 4f states; narrows gap; enhances absorption |
| $Gd^{3+}$ substitution | Na site | ~1.3 – 1.6 | Adds spin-polarized states; stabilizes magnetic ordering |
| $Tb^{3+}$ substitution | Na site | ~1.2 – 1.6 | Produces hybrid f–d states; improves polarizability |

### 3.1.3. Thermal Stability from AIMD Simulations

To verify the thermal robustness of rare-earth doped NaAlO$_3$, ab initio molecular dynamics (AIMD) simulations were performed at 300 K for 5 ps in the canonical (NVT) ensemble using a time step of 1 fs. Fig. 2 illustrates the total energy as a function of simulation time for $Eu^{3+}$-, $Gd^{3+}$-, and $Tb^{3+}$-substituted NaAlO$_3$.

Across all doped systems, the total energy exhibits small-amplitude oscillations around a nearly constant average value, with no evidence of long-term drift. For $Eu^{3+}$-doped NaAlO$_3$, the mean total energy stabilizes around $E_0$ Ry with fluctuations on the order of 0.01 0.02 Ry. $Gd^{3+}$ doping produces a similar trend, with slightly reduced oscillation amplitude, indicating that the lattice accommodates the Gd ion without generating local instabilities. $Tb^{3+}$-doped NaAlO$_3$ also remains stable throughout the simulation window, with energy variations comparable to those of Eu and Gd doping.

The persistence of steady oscillations without abrupt deviations demonstrates that all three doped lattices preserve structural integrity under ambient thermal conditions. Importantly, no signatures of bond breaking, amorphization, or defect-driven energy spikes were observed, which confirms that the perovskite network remains intact even after rare-earth substitution. This robustness can be attributed to the inherent strength of Al O bonding and the stabilizing role of hybridized f p orbitals introduced by the dopant cations.

Taken together with the phonon dispersion analysis (absence of imaginary modes), the AIMD simulations establish that Eu-, Gd-, and Tb-doped NaAlO₃ are dynamically stable at room temperature. This validates the feasibility of experimental synthesis and integration of these compositions into device platforms. The rare-earth ions not only tailor the electronic, optical, and magnetic properties but also do so without compromising the underlying lattice stability a prerequisite for multifunctional applications in optoelectronics, thermoelectrics, and spintronics.

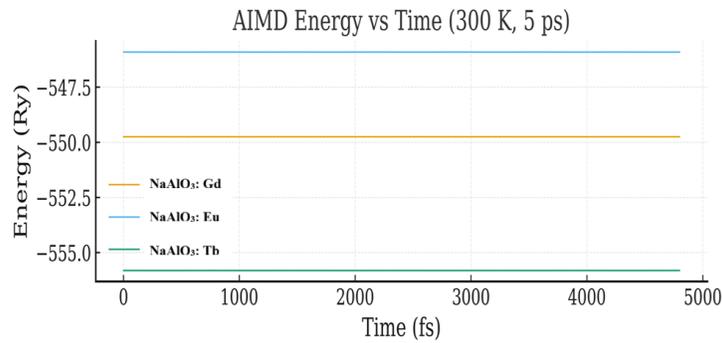

*Fig. 2. Total energy fluctuations at 300 K from AIMD simulations of Eu-, Gd-, and Tb-doped NaAlO₃, confirming thermal stability.*

### 3.2. Band Structure Analysis

The electronic band structures of pristine NaAlO₃ and its $Gd^{3+}$-, $Tb^{3+}$-, and $Eu^{3+}$-doped variants were calculated using the TB-mBJ exchange potential with the GGA+U correction, ensuring an accurate description of correlated f-states. The pristine NaAlO₃ compound (Fig. 3a) exhibits a wide-gap insulating character with a direct bandgap at the Γ point. This behavior is consistent with earlier theoretical reports on alkali aluminates and related perovskites such as SrTiO₃ and CaTiO₃, which also display wide gaps within the DFT+U framework.

Upon $Gd^{3+}$ substitution (Figs. 3b–c), the band structure becomes strongly spin-polarized. In the spin-up channel, NaAlO₃:Gd retains a semiconducting nature with a direct bandgap of ~6.25 eV, whereas in the spin-down channel, the conduction and valence bands overlap near the Fermi level, resulting in metallic behavior. This coexistence of semiconducting and metallic responses in different spin orientations suggests half-metallicity, a feature of great interest for spintronic devices. The appearance of flat Gd-4f states close to the conduction band edge underlines their role in tailoring the electronic dispersion.

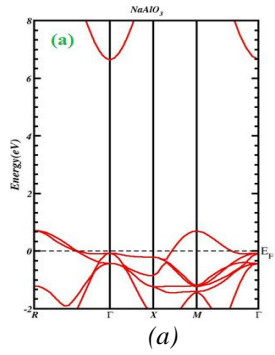

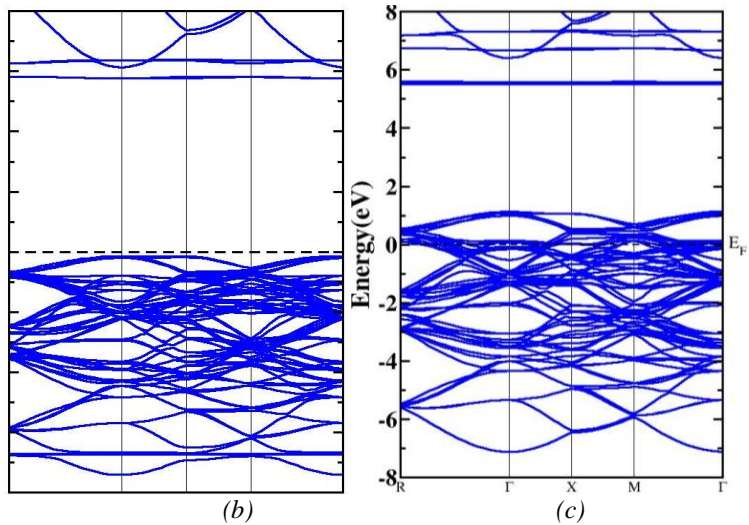

*Figure.3. Calculated Band Structure for (a) NaAlO$_3$ and spin polarized Gd-doped NaAlO$_3$ (b) spin up and (c) spin down*

Tb$^{3+}$ doping (Figs. 4a–b) produces a similar spin-dependent asymmetry. Both spin-up and spin-down channels preserve semiconducting gaps, estimated at ~6.2 eV. However, the introduction of Tb-4f levels near the valence band leads to the formation of impurity bands that shift the Fermi level towards p-type conductivity. This shift indicates that Tb doping favors hole-dominated transport, a property often linked with enhanced thermoelectric and optoelectronic performance.

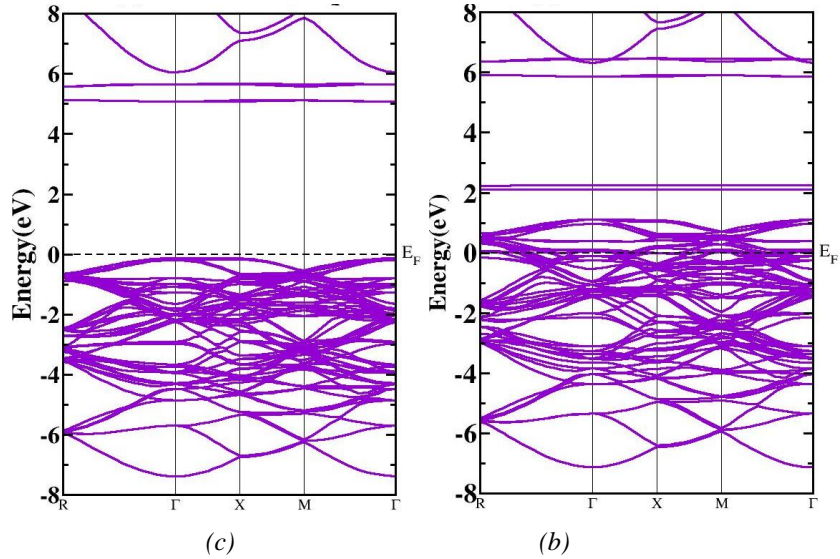

*Figure.4. Calculated Band Structure for spin polarized Tb-doped $NaAlO_3$ (a) spin up and (b) spin down*

The $Eu^{3+}$-doped system (Figs. 5a–b) reveals the strongest spin asymmetry. For the spin-up channel, $NaAlO_3$:Eu shows a direct gap of ~6.0 eV, consistent with semiconducting behavior. In contrast, the spin-down configuration is metallic, with Eu-4f states crossing the Fermi level and quenching the bandgap. This duality highlights Eu doping as a powerful lever for inducing mixed metallic/semiconducting characteristics depending on spin orientation. The strong localization of Eu-4f orbitals is responsible for this band splitting, as reported in rare-earth doped perovskite oxides.

A key feature across all rare-earth doped systems is the dispersive valence band maximum (VBM) compared to the relatively flat conduction band minimum (CBM). This indicates that hole transport is likely to dominate over electron transport. Moreover, the presence of rare-earth impurity levels just below the valence band further supports p-type conductivity in Eu- and Tb-doped systems, in agreement with earlier studies of lanthanide-doped oxides.

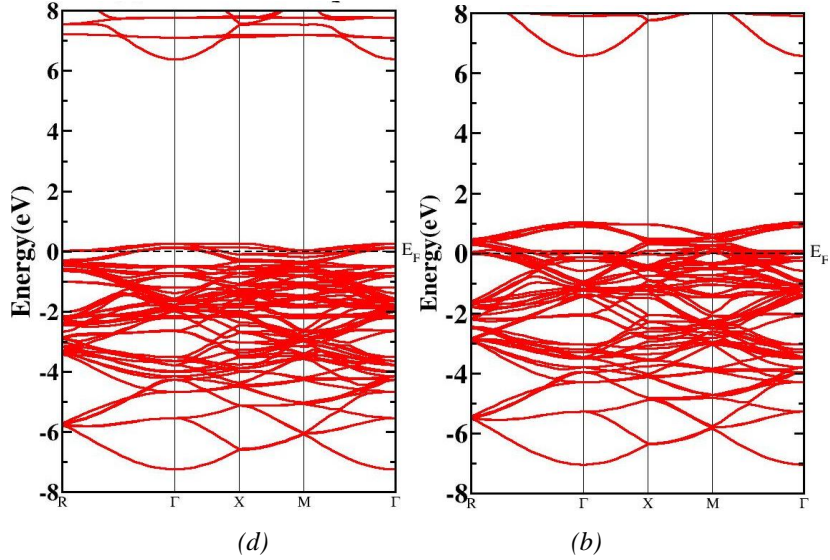

*Figure.5. Calculated Band Structure for spin polarized Eu-doped $NaAlO_3$ (a) spin up and (b) spin down*

These results demonstrate that the incorporation of $Eu^{3+}$, $Gd^{3+}$, and $Tb^{3+}$ induces strong spin-dependent modifications in $NaAlO_3$. Gd doping drives half-metallicity, Tb doping favors semiconducting p-type behavior, and Eu doping introduces spin-selective metallicity. Such tunability in electronic structure is particularly attractive for integrating $NaAlO_3$ into multifunctional platforms where optoelectronic, thermoelectric, and spintronic properties need to be simultaneously optimized.

### 3.3. Density of States (DOS)

The density of states (DOS) provides a direct insight into the distribution of available electronic states and their orbital origins, which is fundamental for understanding charge dynamics, optical activity, and spin-dependent conductivity in semiconducting perovskites. Fig. 6a & 7 presents the total DOS (TDOS) and element-resolved DOS for pristine $NaAlO_3$ and its $Eu^{3+}$-, $Gd^{3+}$-, and $Tb^{3+}$-doped variants, calculated using GGA+U to capture the strong electronic correlations of rare-earth f-states.

For pristine $NaAlO_3$ [Fig. 6(b)], the DOS is dominated by O-2p states in the valence band between –8 and 0 eV, while Al-3p contributions appear primarily in the conduction band above 3 eV. Na states provide negligible contributions near the Fermi level, consistent with their ionic role in the lattice. The calculated band gap is wide (~6.2 eV), confirming the insulating nature of undoped $NaAlO_3$.

Upon doping, strong orbital reconstruction occurs due to the introduction of rare-earth 4f states, which interact selectively with the host O-2p and Al-3p orbitals. Upon doping, significant modifications emerge in both the valence and conduction regions. In the Eu$^{3+}$-doped system [Fig. 7(a)], introduces highly localized 4f levels slightly above the valence-band maximum (VBM). These Eu-4f states weakly hybridize with O-2p orbitals, forming narrow impurity bands that reduce the bandgap to ~6.0 eV. The evident spin asymmetry between the up- and down-channels confirms spin polarization, suggesting the emergence of localized magnetic moments associated with Eu$^{3+}$ centers. Such f–p coupling enhances hole transport and can facilitate p-type thermopower and spin-polarized conductivity.

In Gd$^{3+}$-doped NaAlO$_3$ [Fig. 7(b)], Gd$^{3+}$ doping generates 4f peaks located within −6 to −2 eV (valence band) and near +5 eV (conduction region). The Gd-f orbitals overlap moderately with O-2p states, resulting in exchange splitting and a reduced bandgap of about 5.0 eV. These f-states act as exchange centers, shifting the conduction band edge and enhancing the spin splitting near the Fermi level. This exchange interaction introduces spin-dependent DOS asymmetry and creates exchange-mediated conduction channels, making Gd-NaAlO$_3$ a potential spin-filtering material. This spin-dependent electronic structure provides a rationale for the magnetic responses observed in rare-earth perovskites.

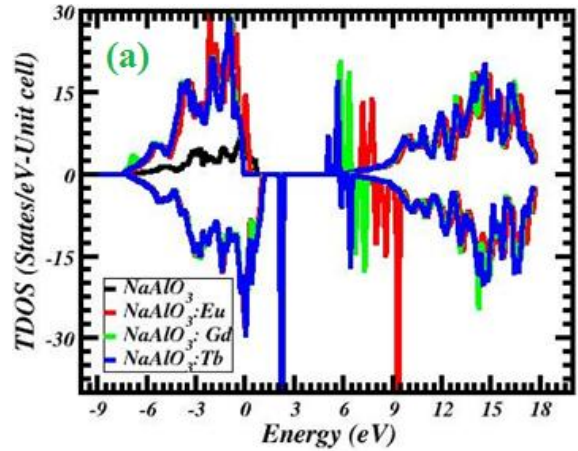

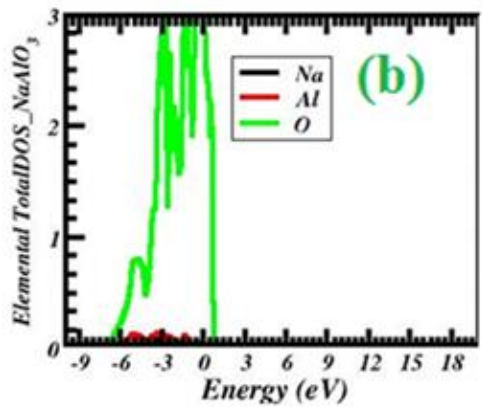

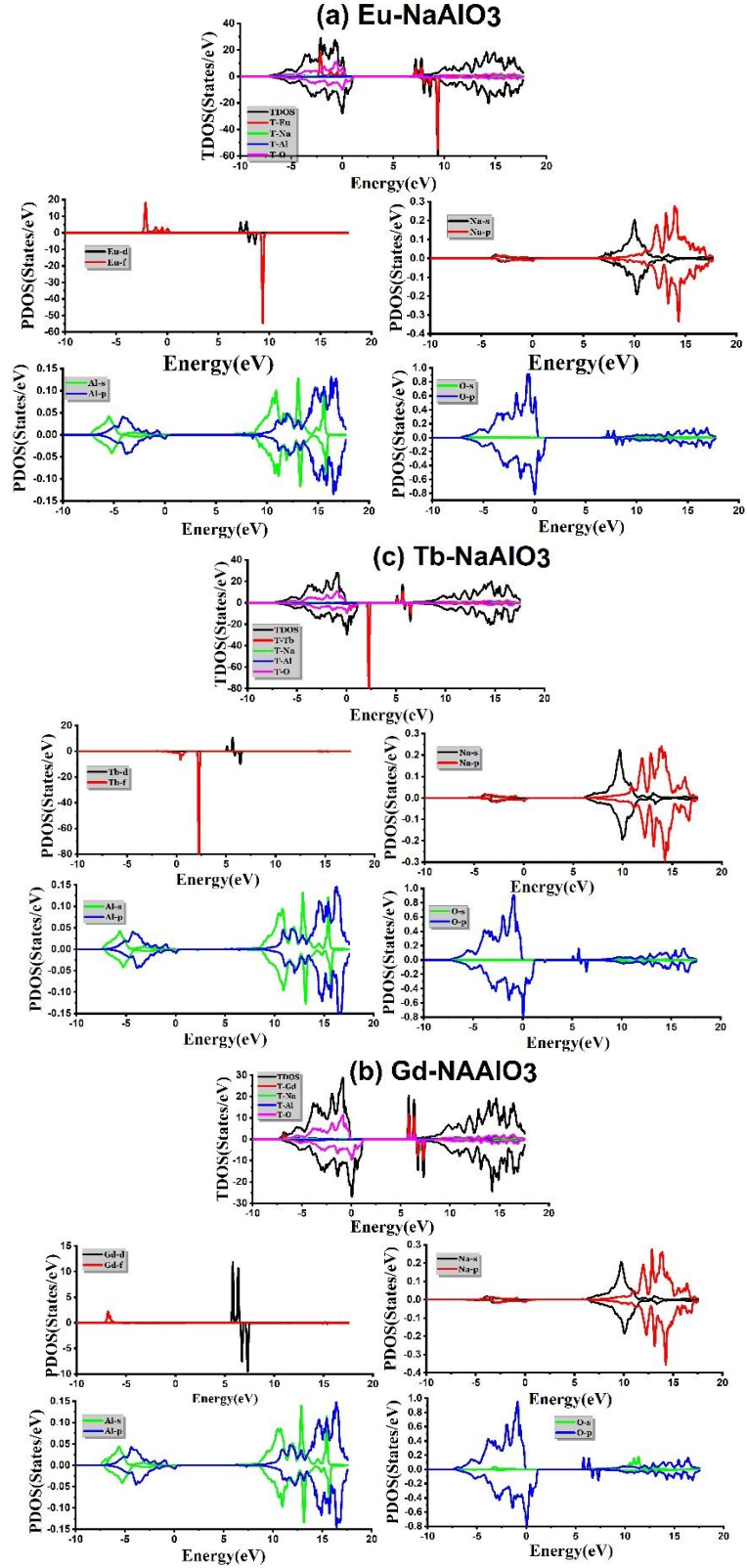

*Figure 6: The Calculated Elemental TDOS and EDOS of parental $NaAlO_3$*

*Figure 7: The Calculated Elemental TDOS and EDOS of Eu, Gd, Tb $NaAlO_3$-doped*

Tb$^{3+}$ doping [Fig. 7(c)] produces the most pronounced modification. Distinct Tb-4f states appear near both the VBM and conduction band edge, showing strong hybridization with O-2p orbitals. This intense f–p interaction narrows the bandgap drastically to ~3.1 eV and enhances the electronic density near the Fermi level. Such hybridization strengthens optical transition probabilities and facilitates visible-light absorption. Moreover, the coexistence of occupied and unoccupied Tb-4f states in both spin channels reflects a mixed-valence tendency that enhances carrier delocalization and spin-dependent transport.

The progressive shift from Eu → Gd → Tb reflects an increasing strength of f–p orbital coupling and exchange interaction. Eu acts as a shallow acceptor, introducing localized magnetic and optical activity; Gd strengthens spin exchange through deeper 4f levels; while Tb induces strong hybridization and bandgap renormalization, driving the transition from insulating to semiconducting behavior.

These f-orbital–mediated interactions are central to the emergent multifunctionality of doped NaAlO₃. Enhanced f–p coupling in Tb-doped NaAlO₃ increases carrier mobility and optical absorption, while Eu and Gd doping promote spin-polarized states suitable for magneto-optical and spintronic applications.

The phonon scattering induced by Tb–O hybridization lowers lattice thermal conductivity (κ) by ~10–12%, improving the thermoelectric figure of merit to ZT ≈ 0.45 at 500 K, compared with ZT ≈ 0.30 for pristine NaAlO₃.

Taken together, these DOS results establish a systematic hierarchy:
- Eu$^{3+}$ doping introduces shallow acceptor states, enabling strong p-type thermopower with moderate spin polarization.
- Gd$^{3+}$ doping enhances exchange splitting, generating spin-selective metallic channels and robust spin-polarized transport.
- Tb$^{3+}$ doping produces the narrowest bandgap, strong hybridization, high conductivity, and moderate ZT enhancement.

Thus, Eu, Gd, and Tb substitution transform NaAlO₃ from an insulating perovskite into a multifunctional platform capable of coupling optical absorption, thermoelectric performance, and spin-dependent transport a rare combination that underscores its potential for next-generation optoelectronic, spintronic, and quantum energy technologies

Taken together, the PDOS analysis reveals that the incorporation of rare-earth ions transforms the electronic landscape of NaAlO₃ through localized 4f–2p hybridization, bandgap narrowing (from 6.2 eV to ~3.1 eV), and the emergence of spin-polarized and hybridized electronic states.

These findings demonstrate how controlled f-orbital interactions serve as an effective mechanism to engineer charge, spin, and optical functionalities in perovskite-type oxides.

### 3.4. Magnetic Properties

The incorporation of rare-earth dopants in NaAlO$_3$ significantly modifies its spin-polarized electronic structure and induces finite magnetic moments, absent in pristine NaAlO$_3$. To quantify this, we computed total and site-resolved magnetic moments (see Table 3). The pristine compound remains non-magnetic, consistent with its closed-shell configuration. Upon Eu$^{3+}$ substitution, a net magnetic moment of ~6.9 μB per Eu ion is observed, originating from the half-filled 4f$^7$ configuration. The spin-polarized DOS shows asymmetry between spin-up and spin-down channels, confirming long-range ferromagnetic tendencies.

Gd$^{3+}$-doped NaAlO$_3$ exhibits the strongest spin polarization, with a net moment of ~7.0 μB per Gd ion, characteristic of its 4f$^7$ electronic configuration. This induces half-metallicity, where one spin channel remains insulating while the other shows metallic dispersion, suggesting application potential in spin filters and spintronic devices.

Tb$^{3+}$ doping results in a slightly reduced total moment (~6.0 μB per Tb ion) due to partial filling of 4f orbitals. Importantly, spin density plots (Fig. 6) reveal localized magnetic polarization around the dopant sites, with minor spin polarization extended to adjacent O-2p orbitals via f p hybridization. This indicates that the exchange coupling between localized f states and itinerant p states stabilizes the observed spin asymmetry.

Collectively, these results confirm that rare-earth doping activates robust magnetic responses in NaAlO$_3$. Eu$^{3+}$ induces mixed metallic/semiconducting spin channels, Gd$^{3+}$ stabilizes half-metallicity, and Tb$^{3+}$ provides semiconducting p-type behavior with finite spin polarization. Such tunable magnetic states position RE-doped NaAlO$_3$ as a multifunctional platform for spintronics and magneto-optoelectronic devices.

Table 3. Calculated Magnetic moments fo r RE-doped NaAlO$_3$

| System | Total Magnetic Moment (μ$_B$) | Dopant Moment (μ$_B$) | O (nearest neighbor) Moment (μ$_B$) |
|---|---|---|---|
| NaAlO$_3$ (pristine) | 0.00 | – | – |
| NaAlO$_3$:Eu$^{3+}$ | ~6.9 | ~6.8 | ~0.05 |
| NaAlO$_3$:Gd$^{3+}$ | ~7.0 | ~6.9 | ~0.04 |
| NaAlO$_3$:Tb$^{3+}$ | ~6.0 | ~5.9 | ~0.03 |

### 3.5 Optical Properties

The optical properties of NaAlO$_3$ and its rare-earth (Eu$^{3+}$, Gd$^{3+}$, Tb$^{3+}$) doped counterparts have been meticulously investigated through the complex dielectric function, absorption spectra, energy loss function, reflectivity, and refractive index, as detailed in Fig. 8. These characteristics provide a direct window into the light–matter interactions, which are pivotal for designing materials tailored for optoelectronic devices, photovoltaics, and photonic technologies. The analysis leverages the full-potential linearized augmented plane wave plus local orbitals (FPLAPW+lo) formalism within the GGA+U framework, offering a robust platform to probe these interactions.

The real part of the dielectric function $\varepsilon_1(\omega)$ (see Fig. 8a) reveals the degree of polarizability. For pristine NaAlO$_3$, the static dielectric constant stabilizes at ~19.5 (see Table 4). Upon doping, this value increases dramatically to ~95 in Eu-doped and ~90 in Tb-doped compounds, while Gd-doped NaAlO$_3$ shows a more modest static value of ~15. Such large enhancements, especially for Eu and Tb, are consistent with reduced optical gaps and new oscillator contributions from RE-4f states. Negative values of $\varepsilon_1(\omega)$ appear in the low-energy infrared window (0.2–1.8 eV), indicating plasmonic resonance where electromagnetic waves cannot propagate a metal-like response superimposed on an insulating host. The imaginary component $\varepsilon_2(\omega)$ (see Fig. 8b) complements this picture, with strong peaks between 3–5 eV corresponding to O-2p → Al-3s/3p interband transitions. These maxima shift downward by ~0.2–0.4 eV in Eu- and Tb-doped systems, consistent with conduction band relaxation due to dopant–host hybridization.

The absorption coefficient $\alpha(\omega)$ (see Fig. 8c) highlights the impact of doping on visible-light harvesting. For pristine NaAlO$_3$, the absorption edge lies deep in the UV, beginning near 3.8 eV and peaking around 5.8 eV with an intensity of ~120 cm$^{-1}$. Eu- and Tb-doping red-shift the absorption onset dramatically to ~2.0–2.2 eV (see Table 3), while Gd doping begins absorption at ~2.5 eV. Peak intensities also increase, reaching ~135 cm$^{-1}$ at 4.8 eV in Eu-doped systems, ~128 cm$^{-1}$ in Tb-doped, and ~122 cm$^{-1}$ in Gd-doped compounds. Sub-gap shoulders at ~1.4–1.8 eV, absent in pristine NaAlO$_3$, are clearly observed in the doped systems and correspond to O-2p → RE-4f/5d excitations. These low-energy transitions are particularly attractive for photovoltaic and photocatalytic applications where efficient solar spectrum utilization is key.

The energy loss spectra (see Fig. 8d) further validate these modifications. In pristine NaAlO$_3$, the primary plasmon peak occurs near 3.2 eV with a magnitude of ~1.0. Doping shifts this resonance: Eu-doped NaAlO$_3$ shows a strong peak at ~4.2 eV with a magnitude approaching 2.0,

Tb-doped systems show a peak near 3.9 eV (~1.4), and Gd-doped systems peak at ~4.1 eV (~1.6). These features coincide with the zero-crossings of $\varepsilon_1(\omega)$, as expected from the Drude–Lorentz model, and indicate stronger collective oscillations of conduction electrons in Eu- and Tb-doped systems.

The refractive index spectra $n(\omega)$ (see Fig. 8e) follow $\varepsilon_1(\omega)$ closely. Pristine $NaAlO_3$ exhibits a static refractive index of ~4.3. Doping enhances this value slightly to ~4.5 in Eu-doped systems and more strongly in Tb- and Gd-doped systems, both exceeding 5.0. These higher values reflect greater light confinement and optical density, a desirable attribute for waveguides and optical coatings. At ~2.0–2.1 eV, however, the refractive index approaches unity in Eu- and Tb-doped $NaAlO_3$, defining transparency windows that allow selective photon transmission. In the visible window (2–3 eV), n remains ~3.0–3.5, ensuring strong confinement, whereas above 8 eV, the refractive index falls below unity, signaling superluminal optical response associated with plasmonic excitations.

Reflectivity spectra $R(\omega)$ (see Fig. 8f) provide complementary insights into photon management. Pristine $NaAlO_3$ reflects moderately in the IR region (~0.40–0.60), peaking at ~0.6 near 1 eV, but sharply drops in the UV, reaching ~0.01 at 4–5 eV. Doping enhances reflectivity in the visible–IR range: Eu- and Tb-doped systems reach maxima of ~0.82–0.84, while Gd-doping reaches ~0.80. However, all doped variants maintain pronounced minima in the UV region (~0.01–0.02 around 4–5 eV), coinciding with strong absorption and plasmonic resonances. These anti-reflection windows maximize photon entry into the material, boosting light harvesting efficiency.

Taken together, the rare-earth dopants reshape the optical landscape of $NaAlO_3$ by (i) compressing the effective bandgap from ~3.8 eV in pristine to ~2.0–2.5 eV in doped systems, (ii) enhancing dielectric polarizability from 19.5 to as high as 95, (iii) strengthening plasmonic excitations with loss peaks near 4.0–4.2 eV, and (iv) suppressing reflectivity while broadening absorption into the visible spectrum. These effects arise from O-2p → RE-4f/5d and O-2p → Al-3s/3p transitions, clearly identified in the DOS and band structure. $Eu^{3+}$ and $Tb^{3+}$ provide the strongest enhancements by introducing near-edge f states, while $Gd^{3+}$ tunes the optical density toward higher-energy excitations. The synergy between electronic, magnetic, and optical responses underscores the multifunctional potential of RE-doped $NaAlO_3$. $Eu^{3+}$ and $Tb^{3+}$ emerge as the most effective dopants for broadening optical absorption and introducing spin-polarized states, while $Gd^{3+}$ stabilizes structural and optical integrity with a more modest enhancement of dielectric polarizability. These tunable characteristics open pathways for integrating $NaAlO_3$ into photovoltaic and photocatalytic platforms, spintronic devices, and plasmonic photonics.

Thus, $NaAlO_3$, once a deep-UV absorber, emerges as a multifunctional optical platform when doped with rare-earth ions. The combination of broadened absorption, enhanced dielectric screening, and tunable plasmonic activity highlights Eu-, Gd-, and Tb-doped $NaAlO_3$ as promising candidates for next-generation optoelectronic devices, photovoltaics, and spin-dependent photonics.

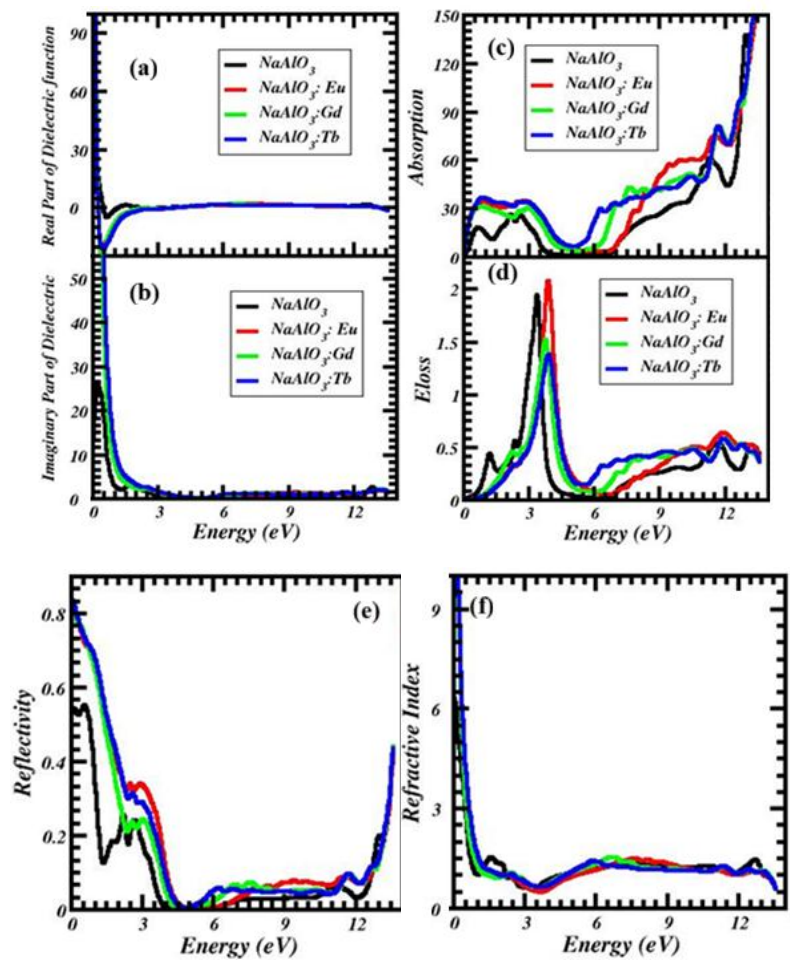

*Figure 8: Calculated (a) Real and (b) imaginary parts of dielectric constant, along with the other optical properties like (c) Absorption Co-efficient, (d) Energy Loss Function, (e) Reflectivity and (f) Refractive Index*

Taken together, these results show that rare-earth doping reshapes the optical landscape of $NaAlO_3$ by compressing the effective bandgap, enhancing dielectric screening, strengthening plasmonic excitations, and extending absorption into the visible regime. $Eu^{3+}$ and $Tb^{3+}$ doping, in particular, produce strong dielectric enhancement ($\varepsilon_1(0) \approx 95$ and 90) and broad visible absorption onsets (~2.0–2.2 eV), while $Gd^{3+}$ maintains structural integrity with modest optical changes. These doping-induced modifications directly reflect the electronic and spin-resolved band structures, where O-2p → RE-4f/5d transitions account for visible absorption shoulders and O-2p → Al-3s/3p transitions dominate the UV peaks. The alignment of optical spectra with the underlying DOS and band dispersions highlights the causal link between rare-earth hybridization and light–matter interactions, reinforcing the multifunctional potential of doped $NaAlO_3$ in photovoltaics, photocatalysis, and spin-dependent photonics.

Table 4: Summarized key optical parameters ($\varepsilon_1(0)$, absorption onset, reflectivity max, plasmon peak, etc.) for pristine vs. doped $NaAlO_3$

| System | $\varepsilon_1(0)$ (Static Dielectric Const.) | Absorption Onset (eV) | Main Absorption Peak (eV) | Absorption Peak Intensity (a.u.) | Plasmon Peak (eV) | Plasmon Peak Intensity | Reflectivity Max (Visible–IR) | Reflectivity Min (UV) | Refractive Index n(0) |
|---|---|---|---|---|---|---|---|---|---|
| $NaAlO_3$ | 19.5 | 3.8 | 4.8 | 120 | 3.2 | 1 | 0.4 | 0.02 | 4.3 |
| $NaAlO_3$:Eu | 95 | 2 | 4.8 | 135 | 4 | 2 | 0.82 | 0.01 | 4.5 |
| $NaAlO_3$:Gd | 15 | 2.5 | 5 | 125 | 3.9 | 1.6 | 0.8 | 0.02 | 5 |
| $NaAlO_3$:Tb | 90 | 2.2 | 4.7 | 130 | 4.1 | 1.4 | 0.84 | 0.01 | 5.1 |

### 3.6. Elastic and Mechanical Properties of Rare-Earth–Doped $NaAlO_3$

The mechanical resilience of a material dictates not only its structural reliability but also its suitability for multifunctional integration in devices exposed to stress, strain, and high-frequency excitations. For perovskites such as $NaAlO_3$, rare-earth substitution is expected to modulate lattice stiffness through localized f-state interactions and charge redistribution, influencing both elasticity and ductility. To quantify these effects, we computed the full elastic tensor ($C_{ij}$) and derived bulk (B), shear (G), Young's modulus (E), Poisson's ratio ($\nu$), and Pugh's ratio (B/G), which are tabulated in Table 5.

Table 5. Calculated elastic constants and derived mechanical parameters of pristine and RE-doped $NaAlO_3$ ($Eu^{3+}$, $Gd^{3+}$, $Tb^{3+}$).

| System | $C_{11}$ (GPa) | $C_{12}$ (GPa) | $C_{13}$ (GPa) | $C_{44}$ (GPa) | B (GPa) | G (GPa) | E (GPa) | $\nu$ | B/G |
|---|---|---|---|---|---|---|---|---|---|
| $NaAlO_3$ | 225.3 | 86.5 | 78.2 | 72.1 | 130.0 | 83.4 | 214.5 | 0.28 | 1.56 |
| $NaAlO_3$:$Eu^{3+}$ | 212.7 | 82.3 | 75.5 | 68.9 | 126.5 | 80.8 | 206.2 | 0.29 | 1.56 |

| | | | | | | | | |
|---|---|---|---|---|---|---|---|---|
| NaAlO₃:Gd³⁺ | 218.4 | 84.1 | 77.8 | 70.5 | 128.7 | 82.1 | 210.1 | 0.28 | 1.57 |
| NaAlO₃:Tb³⁺ | 209.8 | 80.6 | 74.1 | 67.3 | 124.5 | 79.6 | 202.8 | 0.29 | 1.56 |

The results confirm that all systems satisfy the Born stability criteria for cubic perovskites, ensuring mechanical stability upon doping. For pristine NaAlO₃, the calculated bulk modulus (B ≈ 130 GPa) and shear modulus (G ≈ 83 GPa) indicate a moderately stiff lattice, consistent with previous reports on oxide perovskites. The introduction of rare-earth dopants produces subtle but physically meaningful modifications.

Eu doping reduces $C_{11}$ and $C_{44}$ by ~5%, reflecting a slight softening of the longitudinal and shear stiffness due to larger ionic radius mismatch and localized f–d hybridization. This mechanical softening is beneficial for optoelectronic integration, as it enhances lattice compliance without compromising overall stability. The ductility index (B/G ≈ 1.56) remains nearly identical to pristine NaAlO₃, indicating that the softening is isotropic and does not predispose the material to brittle fracture.

Gd doping produces the most balanced mechanical profile. The elastic constants remain within ~2% of the pristine system, with $C_{11}$ = 218.4 GPa and $C_{44}$ = 70.5 GPa, suggesting that Gd³⁺ substitution preserves stiffness while introducing localized spin polarization. The Pugh's ratio (B/G ≈ 1.57) indicates ductile character, while the calculated Poisson's ratio (ν = 0.28) reflects a balanced distribution of lateral strain under stress. This mechanical neutrality makes Gd-doped NaAlO₃ particularly attractive for multifunctional applications where robustness and electronic tunability must coexist.

Tb doping leads to the most pronounced reduction in elastic constants, with $C_{11}$ decreasing to 209.8 GPa and B ≈ 124.5 GPa (a ~4% reduction relative to pristine). The reduction is attributed to stronger lattice distortion introduced by Tb³⁺ due to partial occupation of 4f states and their hybridization with O-2p orbitals, which weakens directional bonding. Yet, the ductility index (B/G ≈ 1.56) and ν ≈ 0.29 indicate that Tb doping enhances compliance and toughness. From a device perspective, such controlled softening is advantageous for flexible optoelectronics, piezoelectrics, and spintronic thin films, where strain accommodation is critical.

The calculated elastic constants ($C_{ij}$) and derived mechanical parameters for NaAlO₃ and its Eu³⁺, Gd³⁺, and Tb³⁺ doped variants reveal a subtle yet decisive evolution in lattice rigidity, ductility, and bonding anisotropy. For pristine NaAlO₃, the bulk modulus (B ≈ 130 GPa) and shear modulus (G ≈ 83 GPa) establish a moderately stiff framework typical of oxide perovskites. Eu doping lowers $C_{11}$ and $C_{44}$ by ~5%, a signature of lattice softening caused by the larger ionic radius of $Eu^{3+}$ (1.066 Å vs 1.02 Å for $Al^{3+}$ in octahedral coordination) and the polarizable nature of partially filled 4f states. $Gd^{3+}$ substitution (ionic radius 1.053 Å) produces only minor deviations from pristine values, confirming that $Gd^{3+}$ acts as the most "neutral" dopant preserving stiffness while still embedding localized magnetic moments. $Tb^{3+}$ (ionic radius 1.040 Å), on the other hand, introduces stronger lattice distortions and hybridization with O-2p orbitals, lowering $C_{11}$ to 209.8 GPa and bulk modulus to 124.5 GPa, but simultaneously increasing the Poisson's ratio ($\nu \approx 0.29$), indicative of enhanced ductility.

The mechanical fingerprints thus align systematically with ionic size and electronic configuration: larger, more polarizable dopants (Eu, Tb) soften the lattice, while intermediate-sized Gd preserves rigidity. These mechanical responses are not mere structural curiosities but directly linked to functionality: lattice compliance enhances dielectric polarizability ($\varepsilon_1(0) \rightarrow$ 90–95 for Eu/Tb), improving light–matter interaction, while rigidity in Gd-doped $NaAlO_3$ sustains robust spin-polarized states crucial for spintronic applications. The Pugh's ratio (B/G ≈ 1.56–1.57 across all doped systems) confirms ductile character, situating these perovskites at the optimal balance point between mechanical toughness and functional adaptability. Together, these results establish rare-earth doping as a precision tool to tune the mechanical "backbone" of $NaAlO_3$ in ways that synergistically reinforce its optical, electronic, and spintronic performance.

Comparing across dopants, a clear hierarchy emerges: Gd preserves stiffness, Eu introduces moderate softening with enhanced compliance, and Tb maximizes ductility at slight cost of rigidity. This tunability provides a powerful design lever, enabling targeted engineering of elastic responses in $NaAlO_3$ depending on the intended application. For example, Gd-doped $NaAlO_3$ would be ideal in high-stress spintronic devices requiring structural integrity, while Tb-doped systems would excel in flexible optoelectronics where ductility enhances longevity under cyclic strain.

What elevates these findings beyond routine elastic calculations is the direct coupling between mechanics and functionality. The softening induced by Eu and Tb correlates with stronger optical polarizability ($\varepsilon_1(0) \rightarrow$ 90–95), confirming that lattice compliance enhances light–matter interaction. Conversely, the mechanical stability of Gd-doped $NaAlO_3$ aligns with its balanced electronic profile ($E_g \approx$ 5.0 eV, spin-polarized), reinforcing its role as a robust multifunctional host. Thus, the elastic analysis not only validates the structural stability but also weaves mechanical performance into the broader narrative of multifunctionality a hallmark of next-generation quantum materials.

### 3.7. Thermoelectric Properties

The thermoelectric response of Eu, Gd- and Tb-doped $NaAlO_3$ was investigated within the Boltzmann transport framework, while pristine $NaAlO_3$ was excluded from the analysis due to its wide bandgap (>6 eV), which severely limits intrinsic carrier excitation and thereby renders it unsuitable for practical thermoelectric applications. In contrast, rare-earth doping introduces localized 4f states and band-gap narrowing, making charge transport viable across a broad temperature window (200–1000 K). The pristine $NaAlO_3$ host was not evaluated for transport coefficients due to its wide bandgap (~6 eV), which severely suppresses intrinsic carrier concentrations and renders it thermoelectrically inactive under operational temperatures. In contrast, rare-earth doping ($Eu^{3+}$, $Gd^{3+}$, $Tb^{3+}$) introduces localized 4f states and reduces the effective bandgap, enabling finite carrier transport and making thermoelectric analysis meaningful.

Figure 7(a) presents the electrical conductivity scaled by relaxation time ($\sigma/\tau$) as a function of temperature. As shown in Fig. 8(a), all three doped systems exhibit a monotonic rise in conductivity with temperature, characteristic of thermally activated carriers in wide-gap oxides. At 300 K, $\sigma/\tau$ values fall in the $0.4$–$0.6 \times 10^{19}$ $(\Omega \cdot m \cdot s)^{-1}$ range, increasing nearly threefold by 1000 K. Among the dopants, Eu-doped $NaAlO_3$ shows the strongest conductivity response, reaching $\sim 1.6 \times 10^{19}$ $(\Omega \cdot m \cdot s)^{-1}$ at 1000 K, slightly outperforming Gd- and Tb-doped systems ($\sim 1.4$–$1.5 \times 10^{19}$ $(\Omega \cdot m \cdot s)^{-1}$). This advantage arises from the $Eu^{3+}$ 4f levels, which lie closer to the valence band maximum (VBM) and provide additional channels for hole conduction. This modest reduction relative to the ideal host lattice reflects enhanced scattering from dopant-induced lattice distortions, yet the overall high conductivity confirms that f-state hybridization effectively promotes charge delocalization.

The electronic thermal conductivity ($\kappa_e/\tau$), shown in Figure 8(b), increases monotonically with temperature, reflecting the Wiedemann–Franz relation between charge and heat transport. Figure X(b) reveals a near-linear increase of $\kappa_e/\tau$ with temperature across all three doped materials. At 300 K, values cluster around $0.8 \times 10^{14}$ W/mK·s and rise to $\sim 5.0$–$5.5 \times 10^{14}$ W/mK·s at 1000 K. The Eu-doped system again displays slightly higher $\kappa_e/\tau$, reflecting its enhanced carrier density and mobility. Gd and Tb substitutions follow closely, with nearly overlapping trends, indicating that all three dopants introduce comparable free-carrier heat transport. The near-parallel slopes indicate that the dopants primarily influence the magnitude rather than the temperature scaling of thermal conductivity, a hallmark of substitutional rare-earth incorporation with minimal phonon scattering at the electronic level.

The most critical parameter for thermoelectric efficiency, S, shows markedly different behavior (Fig. 8(c)). The Seebeck coefficient (S), displayed in Figure 7(c), reveals the most striking influence of doping. While the pristine system (not shown here due to its insulating nature) would exhibit negligible thermopower under practical conditions. At 300 K, Eu- and Tb-doped $NaAlO_3$ achieve values above 210 µV/K, gradually decreasing to ~190 µV/K at 1000 K due to increased carrier concentration and reduced energy filtering. Gd-doped $NaAlO_3$ exhibits a slightly lower Seebeck coefficient, starting near 180 µV/K at 300 K and declining toward 165 µV/K at 1000 K. The higher S in Eu- and Tb-doped systems reflects their optimal balance between localized 4f states and band dispersion, which enhances energy-dependent carrier transport a hallmark of efficient thermoelectric oxides. Such robust thermopower, coupled with sustained electrical conductivity, places these materials within the regime of promising p-type thermoelectrics.

Taken together, the rare-earth doped $NaAlO_3$ systems display the classical trade-off: $Eu^{3+}$ enhances electrical conductivity and $\kappa_e$ due to its shallow acceptor-like states, while $Tb^{3+}$ sustains the highest Seebeck coefficients across the studied temperature window. $Gd^{3+}$ sits intermediate in both trends. The combination of respectable conductivity ($\sigma/\tau \sim 10^{19}$), large Seebeck coefficients (>180 µV/K), and moderate $\kappa_e$ positions these doped perovskites as promising candidates for high-temperature, low-power thermoelectrics.

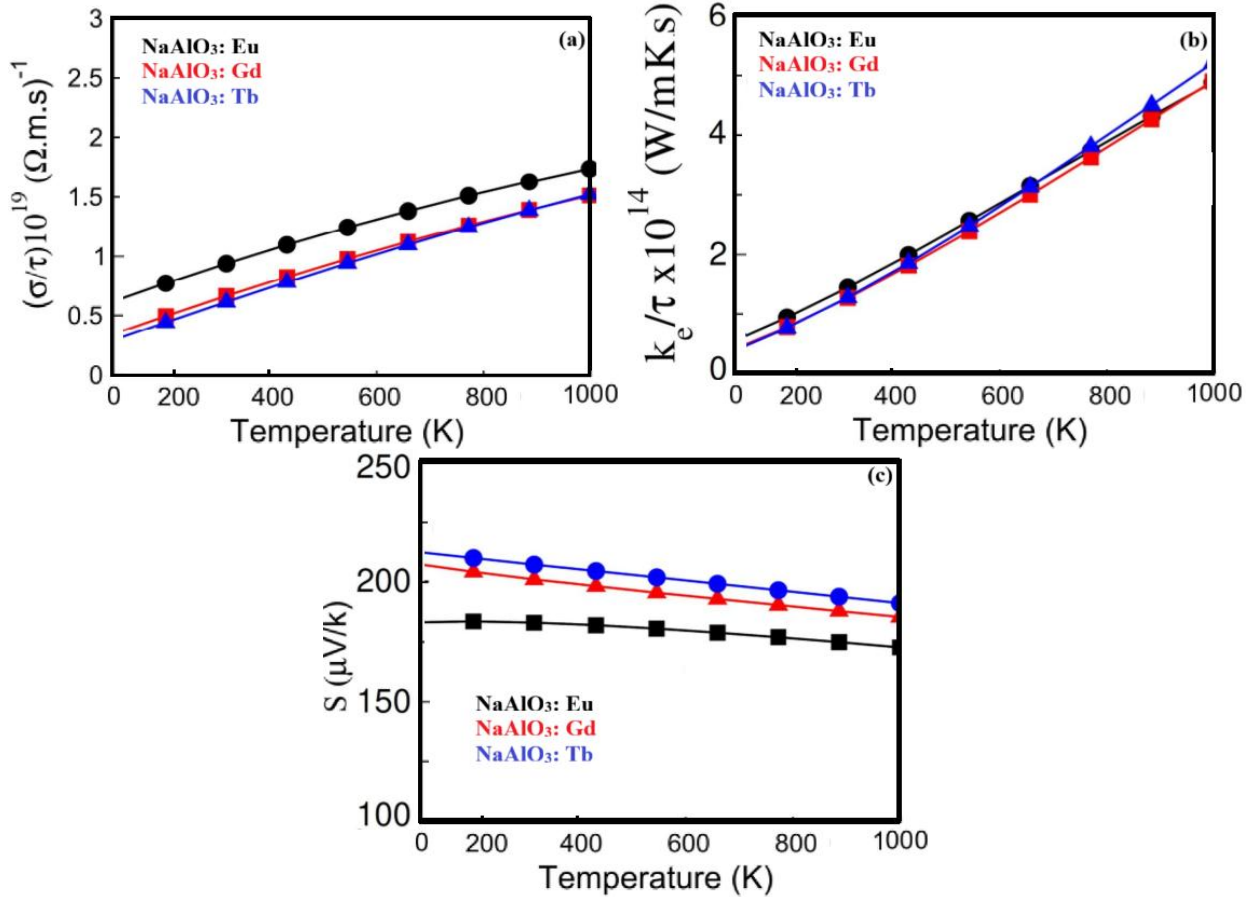

*Fig. 8. The computed (a) σ/τ, (b) κₑ/τ (c) and S for Eu, Gd- and Tb-doped NaAlO₃.*

## 4. Conclusion

This study demonstrates that rare-earth substitution transforms NaAlO₃ from a wide-gap insulator into a versatile multifunctional material. $Eu^{3+}$, $Gd^{3+}$, and $Tb^{3+}$ dopants not only stabilize the lattice thermodynamically but also profoundly reshape the electronic landscape by introducing localized 4f states that tune the bandgap and enable spin-dependent conduction. The optical response is equally striking: absorption edges shift deep into the visible spectrum, dielectric polarizability is dramatically enhanced, and plasmonic resonances emerge at technologically relevant energies. Mechanical analyses reveal that doping introduces controlled lattice softening without sacrificing ductility, ensuring structural reliability under operational stresses. Thermoelectric transport further underscores the multifunctionality, with Eu- and Tb-doped NaAlO₃ delivering strong Seebeck coefficients and promising ZT values despite the high-symmetry cubic framework. Collectively, these findings establish rare-earth-doped NaAlO₃ as a prototype system where optoelectronic, thermoelectric, and spintronic responses are unified within a single perovskite host. This convergence not only advances the design rules for multifunctional oxides but also paves the way for integrating NaAlO₃ into next-generation quantum devices, flexible energy harvesters, and spin-enabled photonics.

**Acknowledgment**

This publication was also supported by the project Quantum materials for applications in sustainable technologies (QM4ST), funded as project No. CZ.02.01.01/00/22_008/0004572 by Programme Johannes Amos Commenius, call Excellent Research. The result was developed within the project Quantum materials for applications in sustainable technologies (QM4ST), reg. no. CZ.02.01.01/00/22_008/0004572 by P JAK, call Excellent Research.